\documentclass [twocolumn, aps, prb, superscriptaddress, showpacs] {revtex4}
\usepackage{graphicx,latexsym,makeidx,amsmath}
\begin{document}
\draft
\title {Quantum Hall Effect in Bernal Stacked and Twisted Bilayer Graphene Grown on Cu by Chemical Vapor Deposition}
\author {Babak Fallahazad}
\affiliation {Microelectronics Research Center, The University of
Texas at Austin, Austin, TX 78758}
\author {Yufeng Hao}
\affiliation {Department of Mechanical Engineering and the Materials Science and Engineering Program,
The University of Texas, Austin, TX 78712}
\author {Kayoung Lee}
\affiliation {Microelectronics Research Center, The University of
Texas at Austin, Austin, TX 78758}
\author {Seyoung Kim}
\affiliation {Microelectronics Research Center, The University of
Texas at Austin, Austin, TX 78758}
\author {R. S. Ruoff}
\affiliation {Department of Mechanical Engineering and the Materials Science and Engineering Program,
The University of Texas, Austin, TX 78712}
\author {E. Tutuc}
\affiliation {Microelectronics Research Center, The University of
Texas at Austin, Austin, TX 78758}
\date{\today}
\begin{abstract}
We examine the quantum Hall effect in bilayer graphene grown on Cu substrates by chemical vapor deposition. Spatially resolved Raman spectroscopy
suggests a mixture of Bernal (A-B) stacked and rotationally faulted (twisted) domains. Magnetotransport measurements performed on bilayer domains
with a wide 2D band reveal quantum Hall states (QHSs) at filling factors $\nu=4, 8, 12$ consistent with a Bernal stacked bilayer, while magnetotransport measurements in bilayer domains defined by a narrow 2D band show a superposition of QHSs of two independent monolayers. The analysis of the Shubnikov-de Haas oscillations
measured in twisted graphene bilayers provides the carrier density in each layer as a function of the gate bias and the inter-layer capacitance.
\end{abstract}
\maketitle

Bilayer graphene consisting of two closely spaced graphene monolayers are an interesting electron system. If the two graphene monolayers forming the bilayer
are Bernal stacked, the system possesses a tunable energy band-gap\cite{ohta,castro,oostinga,zhang}, which renders it attractive
for electronic and optoelectronic applications. While electron transport in natural bilayer graphene has been explored to a large extent,
much less is known about the transport properties of {\it grown} graphene bilayers. Recent studies have reported the growth of bilayer
graphene on SiC and metal substrates by chemical vapor deposition (CVD). Bilayer graphene grown on SiC substrates has been shown to be
Bernal stacked when grown on the Si-face \cite{riedl,lee2011}, and rotationally twisted when grown on the C-face.\cite{hicks} While several
recent studies suggest the growth of Bernal stacked bilayer on metal substrates based on Raman spectroscopy, \cite{lee2010,yan,yan2,luo}
evidence of stacking from electron transport data in grown bilayer graphene has been scant. It is therefore interesting
to probe the electronic properties of CVD-grown graphene bilayers, which in turn can shed light on the growth mechanism
and help assess its potential for applications. Here we provide a systematic investigation of the quantum Hall effect
in bilayer graphene grown on Cu substrates by chemical vapor deposition. Our data show that such bilayers consist
of a {\it mixture} of domains which are either Bernal stacked or are rotationally faulted ('twisted bilayer').

The graphene samples studied here are grown on a 25 $\mu$m-thick Cu foil at a temperature of 1035 $^{\circ}$C by CVD,
using a mixture of methane and hydrogen at the partial pressures of 0.02 mbar and 0.03 mbar, respectively. After the growth,
the graphene film on one side of the Cu foil is coated with PMMA and placed in an aqueous solution of ammonia persulfate ((NH$_4$)$_2$S$_2$O$_8$)
to dissolve the Cu on the unprotected side. The PMMA film that carries the graphene flake is rinsed several times with de-ionized water
to minimize the chemical contamination, and then transferred onto a silicon substrate covered with 285 nm-thick thermally grown SiO$_2$.
After the transfer the sample is allowed to dry, and the PMMA is dissolved in acetone.

\begin{figure*}
\centering
\includegraphics[scale=0.58]{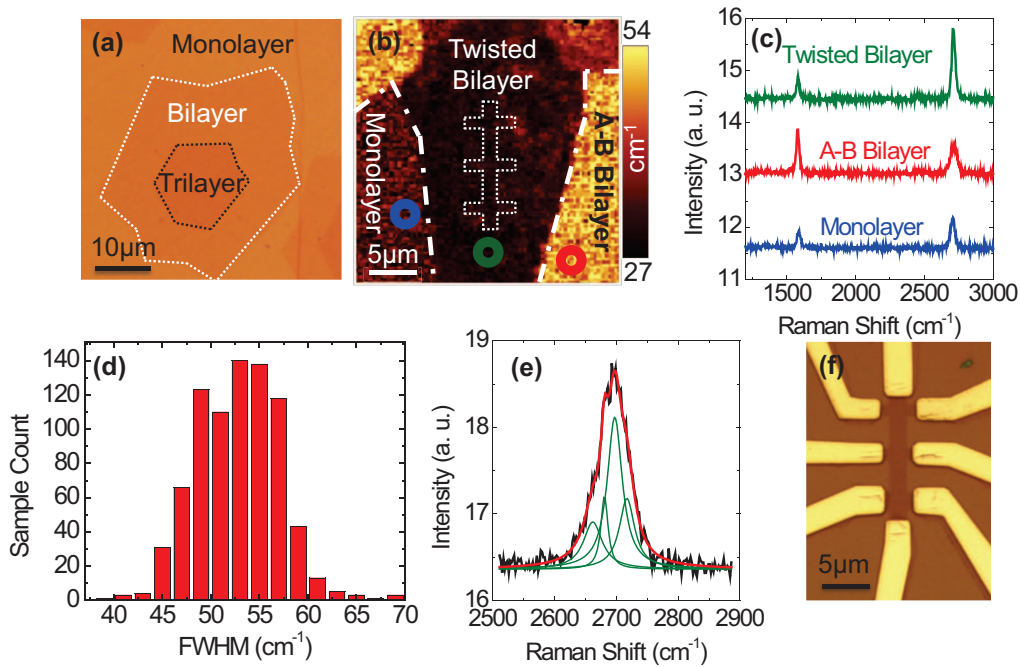}
\caption {\small{(a) Optical micrograph of a multi-layer graphene sample after transfer onto a 285-nm thick SiO$_2$/Si substrate.
The dashed lines delineate monolayer, bilayer, and trilayer regions. (b) 2D FWHM spatial map reveals the bilayer is a mixture of domains with
either wide ($45-54$ cm$^{-1}$) or narrow ($27-33$ cm$^{-1}$) 2D band. The dotted line marks a Hall bar subsequently fabricated to probe electron
transport in individual bilayer regions. (c) Raman spectra acquired at three different positions, as marked in panel (b) show the G ($\simeq1580$ cm$^{-1}$)
and 2D ($\simeq2700$ cm$^{-1}$) bands. (d) Histogram of the 2D band FWHM on a bilayer domain with wide 2D band. The average 2D FWHM is $53\pm2$ cm$^{-1}$.
(e) Example of a 2D band spectrum (black line) acquired on a Bernal stacked bilayer domain. A fit (red) using four Lorentzian functions (green)
provides a very good match to the experimental data. (f) Optical micrograph of a back-gated Hall bar fabricated on bilayer graphene.}}
\end{figure*}

Figure 1(a) shows an optical micrograph of the graphene film transferred on the SiO$_2$ substrate, indicating the presence of monolayer,
bilayer, and trilayer regions. To probe the number of graphene layers, and obtain an initial assessment of the layer
stacking, the sample is characterized by Raman spectroscopy acquired using a 488 nm excitation wavelength, 300 nm spot size, and a power
lower than 0.1 mW. Figure 1(b) presents a representative mapping of the Raman 2D band full width at half maximum (FWHM) acquired over a
30$\times$30 $\mu$m$^2$ area. These data reveal the presence of distinct domains on the bilayer area with either a narrow 2D band, with FWHM values
between 27 and 33 cm$^{-1}$, or a wide 2D band, with FWHM values between 45 and 54 cm$^{-1}$. By comparison the Raman 2D FWHM measured
in monolayer graphene is $28-30$ cm$^{-1}$, while in Bernal (A-B) stacked bilayer graphene the 2D band is wider.\cite{ferrari,casiraghi,ferrari2,malard,hao}
Figure 1(b) data therefore suggest that the bilayer domains with narrow 2D band consist of two graphene monolayers which are rotationally
faulted (twisted bilayer), while the domains characterized by wider 2D band consist of two Bernal stacked monolayers. We note the two types
of bilayer domains of Fig. 1(b) show no obvious differences in optical contrast. Figure 1(c) shows samples of Raman spectra acquired on
the same sample of Fig. 1(b), at different positions on the monolayer, the twisted bilayer, and the Bernal stacked bilayer regions, as indicated.
The 2D FWHM of these Raman spectra are 28 cm$^{-1}$, 30 cm$^{-1}$, and 50 cm$^{-1}$, respectively. The 2D band intensity ($I_{2D}$)
is larger than the G band intensity ($I_G$) on the monolayer and bilayer domains with narrow 2D band, an observation which agrees with Raman
spectroscopy results for exfoliated graphene.\cite{ferrari,casiraghi} In contrast, the bilayer domain with a wide 2D band shows an
$I_{2D}$/$I_G$ ratio lower than 1. The D band, located at a Raman shift of 1350 cm$^{-1}$ is either absent or very weak,
indicating that the defect density is low.

Figure 1(d) shows a histogram of the 2D band FWHM values acquired over a 15$\times$20 $\mu$m$^2$ bilayer graphene grain characterized by
a wide 2D band. The data points range between 45 cm$^{-1}$ and 65 cm$^{-1}$, with a maximum at 53 cm$^{-1}$. Figure 1(e) presents a
typical spectrum of the 2D band selected from the bilayer graphene region with a wide 2D band. Figure 1(e) data
could not be fitted well using a single Lorentzian, but an excellent fit is obtained using four Lorentzian functions.
The combined data of Fig. 1(b-e) therefore suggest that bilayer domains with narrow 2D band consist of twisted graphene monolayers,
while the bilayer domains with wide 2D band are two Bernal stacked monolayers. We next focus on the magneto-transport
in these two types of bilayers.

After the graphene is characterized by Raman spectroscopy, we fabricate back-gated Hall bar devices on selected bilayer domains with
a uniform 2D peak FWHM, which is either narrow ($27-33$ cm$^{-1}$) or wide ($45-65$ cm$^{-1}$). The active region of the Hall bar is
defined by electron-beam (e-beam) lithography and isolated from the rest of the film using oxygen plasma etching. Metal (Ni) contacts
are defined by a second e-beam lithography, metal deposition, and lift-off [Fig. 1(f)]. The carrier mobility ($\mu$) of each sample is determined from the four-point conductivity ($\sigma$) dependence on back-gate bias ($V_{BG}$), $\mu=1/C_{BG}\times d\sigma/dV_{BG}$; $C_{BG}$ is the back-gate capacitance per unit area. For the samples examined in this study $C_{BG} $lies in the range $12.5 - 14.4$ nF$\cdot$cm$^{-2}$, values measured on metal pads deposited in proximity of the Hall bars,
and by Hall measurements. The extracted mobility of the Bernal stacked bilayer graphene devices range between $700-1,800$ cm$^2\cdot$V$^{-1}\cdot$s$^{-1}$
at room temperature and $1,800-2,200$ cm$^2\cdot$V$^{-1}\cdot$s$^{-1}$ at 0.3 K. The twisted bilayers exhibit mobility values between $3,400-3,700$ cm$^2\cdot$V$^{-1}\cdot$s$^{-1}$ at room temperature and $5,300-7,600$ cm$^2\cdot$V$^{-1}\cdot$s$^{-1}$ at 0.3 K. The higher mobility in twisted bilayers by comparison to Bernal stacked bilayers can be explained by differences in their band-structure, which forbid electron back-scattering in monolayer graphene,
and hence in twisted bilayer, but allow back-scattering in Bernal stacked bilyers.\cite{adam}

\begin{figure*}
\centering
\includegraphics[scale=0.72]{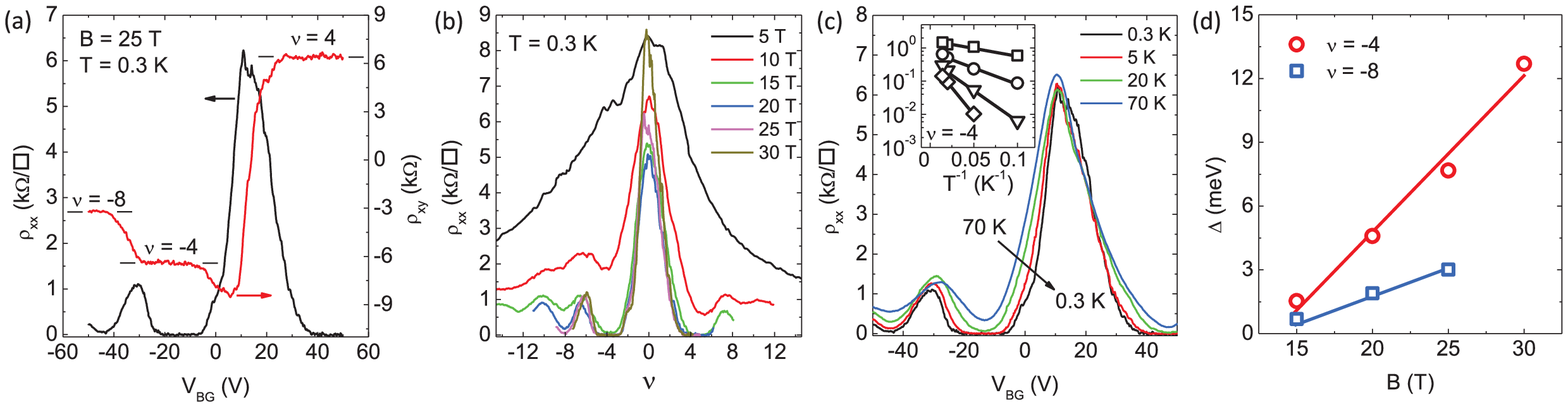}
\caption {\small{(a) $\rho_{xx}$ and $\rho_{xy}$ vs. $V_{BG}$, measured at $B=25$ T, and $T=0.3$ K. The data shows QHSs,
marked by vanishing $\rho_{xx}$ and quantized $\rho_{xy}$, at filling factors $\nu=\pm4$ and $\nu=-8$.
(b) $\rho_{xx}$ vs. $\nu$ measured at $T=0.3$ K, and at different $B$-field values, illustrating the emergence of QHSs
at integer filling factors that are multiples of four with increasing the $B$-field. (c) $\rho_{xx}$ vs. $V_{BG}$ measured at different temperatures,
and at $B=25$ T. Inset: $\rho_{xx}$ vs. $T^{-1}$ at $\nu=-4$ on a log-lin scale, measured at $B=15$ T ($\Box$), 20 T ({\large $\circ$}), 25 T ({\small $\bigtriangledown$}), 30 T ({\large $\diamond$}). (d) $\Delta$ vs. $B$, for $\nu=-4$ and $\nu=-8$ QHSs. The solid lines are guide to the eye.}}
\end{figure*}

To establish the layer stacking of the CVD grown graphene bilayers and explore their electronic properties, in the following we discuss
the quantum Hall effect in this system. Magnetotransport measurements were carried out in perpendicular magnetic fields ($B$)
up to 31 T, using a pumped $^{3}$He refrigerator with a base temperature $T=0.3$ K, and small signal, low frequency lock-in techniques.
Figure 2(a) shows the longitudinal ($\rho_{xx}$) and Hall resistivity ($\rho_{xy}$) as a function of $V_{BG}$ measured at a perpendicular
magnetic field $B=25$ T and $T=0.3$ K in a graphene bilayer that displays a Raman signature consistent with Bernal stacking, i.e. wide 2D band.
The data show clear quantum Hall states (QHSs), marked by vanishing $\rho_{xx}$ and quantized $\rho_{xy}$ at filling factors $\nu=\pm4$ and $\nu=-8$.
The filling factors are determined from the $\rho_{xy}$ plateau values, which are equal to $h/\nu e^2$; $h$ is Planck's constant and $e$
the electron charge. Alternatively, the filling factor can be calculated using $\nu=nh/eB$, where $n$ is the total carrier density calculated
by $n=C_{BG}(V_{BG}-V_D)/e$; $V_D$ is the gate bias at the charge neutrality (Dirac) point.

Figure 2(b) shows the $\rho_{xx}$ vs. $\nu$ measured in the same sample at different $B$ values, and at $T=0.3$ K. The data show the emergence
of QHSs at integer filling factors that are multiples of four, i.e. $\nu=\pm4,-8,-12$, thanks to the fourfold degeneracy of each Landau level (LL)
associated with the spin and valley degrees of freedom.\cite{mccann} The QHSs filling factors of Fig. 2(a,b) are consistent with the expected values
in natural bilayer graphene,\cite{mccann,novoselov} in effect fingerprinting a Bernal stacked bilayer. Figure 2(c) shows $\rho_{xx}$ vs. $V_{BG}$ measured
at $B=25$ T, and at different temperatures. Although the $\nu=\pm4,-8$ QHSs weaken with increasing $T$, these QHSs remain
clearly visible at the highest temperature, $T=70$ K. The inset of Fig. 2(c) shows the Arrhenius plot of
$\rho_{xx}$ measured at $\nu=-4$, and at $B=15, 20, 25, 30$ T. These data follow a thermally activated behavior,
$\rho_{xx}\propto e^{-\Delta/(2k_B T)}$, where $\Delta$ is the energy gap and $k_B$ is Boltzmann's constant.
Figure 2(d) shows the extracted $\nu=-4,-8$ QHSs energy gaps vs. $B$. The data follow a linear dependence of
$\Delta$ as a function of $B$, with the $\Delta$ values approaching 0 at $B\simeq13$ T.
The QHS energy gaps of Fig. 2(d) are considerably smaller than theoretical values.\cite{milton}
For example, the theoretically expected energy gap of $\nu=-4$ at $B=30$ T is 108 meV, a value roughly eight times larger than the experimental value.
The $\nu=-4$ and $\nu=-8$ QHSs energy gaps probed in CVD-grown bilayer graphene are also approximately fivefold smaller than
values typically measured in exfoliated bilayer graphene on SiO$_2$ substrates.\cite{zeitler,kurganova}

\begin{figure*}
\centering
\includegraphics[scale=0.6]{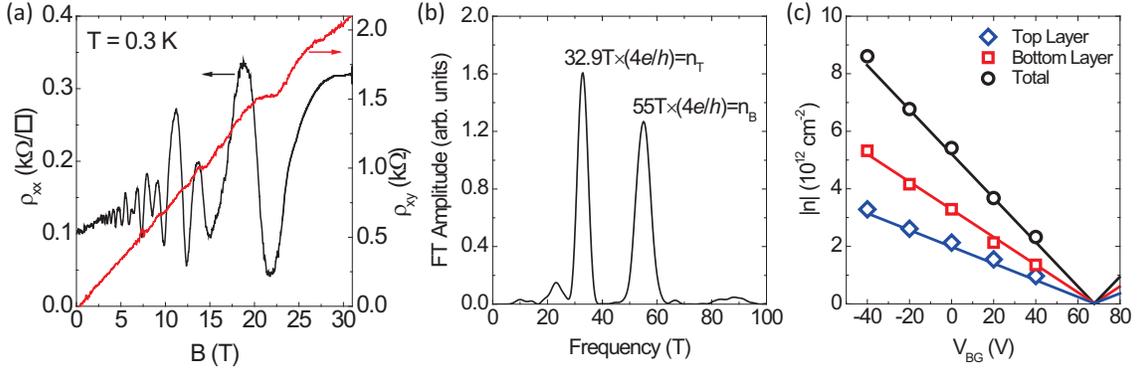}
\caption {\small{(a) $\rho_{xx}$ and $\rho_{xy}$ vs. $B$ measured at $n=-8.4\times10^{12}$ cm$^{-2}$ and at $T=0.3$ K.
The SdH oscillations stem from a QHSs superposition of the two decoupled graphene monolayers.
(b) Fourier transform of $\rho_{xx}$ vs. $B^{-1}$ data. The two peaks represent the layer densities, up to a factor $4e/h$.
(c) Top layer, bottom layer, and total carrier densities of the twisted bilayer graphene vs. $V_{BG}$. The symbols (lines)
represent experimental data (calculations).}}
\end{figure*}

We now turn to the magneto-transport properties of the twisted bilayer graphene samples, fabricated on bilayer graphene domains
with a narrow Raman 2D band. Figure 3(a) shows an example of $\rho_{xx}$ and $\rho_{xy}$ vs. $B$ data, measured in a twisted bilayer device
at $V_{BG}=-40$ V, corresponding to $n=-8.4\times10^{12}$ cm$^{-2}$, and at $T=0.3$ K; the sample mobility is $\mu=7,400$ cm$^2\cdot$V$^{-1}\cdot$s$^{-1}$.
These data possess several noteworthy features. First, the $\rho_{xx}$ vs. $B$ data display Shubnikov-de Haas (SdH) oscillations present
down to magnetic fields as low as $B\simeq3$ T, which contrast Fig. 2(b) data, where QHSs are not visible at $B$-fields lower than 10 T.
This observation can be explained by the larger monolayer graphene LL energies by comparison to bilayer graphene. Moreover,
the $\rho_{xx}$ vs. $B$ data do not follow a QHS sequence which can be readily attributed to either monolayer ($\nu=\pm2,6,10...$)
or bilayer ($\nu=\pm4,8,12...$), and instead shows a beating pattern similar to the QHSs superposition of a multisubband system,
consistent with parallel electron transport in two independent graphene monolayers.

\begin{figure}
\centering
\includegraphics[scale=0.5]{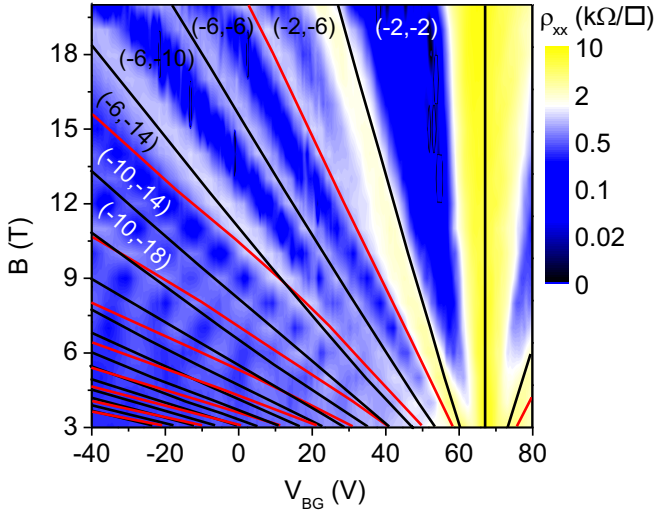}
\caption {\small{Twisted bilayer $\rho_{xx}$ contour plot as a function of $V_{BG}$ and $B$. The red (black) lines
are the calculated position of $\nu=\pm0,4,8,12...$ of the top (bottom) layer. The layer filling factors $(\nu_{T},\nu_{B})$
are indicated for each QHS.}}
\end{figure}

To determine the subband (layer) density in a twisted bilayer, we examined the Fourier transform (FT) of $\rho_{xx}$ vs. $B^{-1}$ data, calculated by first
re-plotting the $\rho_{xx}$ vs. $B$ data, subtracting a linear fit background to center the $\rho_{xx}$ vs. $B^{-1}$ data around zero, and then
applying a fast Fourier transform algorithm. Figure 3(b) shows the FT amplitude vs. $B$ corresponding to Fig. 3(a) data. These data show two
prominent peaks, which yield the two layer densities, up to a factor $4(e/h)=9.67\times10^{10}$ cm$^{-2}\cdot$T$^{-1}$. Figure 3(c) summarizes the
layer and total densities as a function of $V_{BG}$. We attribute the higher (lower) density to the bottom (top) layer, as it lies closer (farther)
with respect to the back-gate. Both layer densities go to zero at $V_D=68$ V.

To understand the top ($n_{T}$) and bottom ($n_{B}$) layer density dependence on $V_{BG}$ in twisted bilayer graphene, we employ a model used to calculate the layer densities in independently contacted graphene double layers separated by a dielectric.\cite{kim} The applied $V_{BG}$ is distributed partly across
the SiO$_2$ dielectric and partly on the Fermi energy of the bottom layer:
\begin{equation}
e(V_{BG}- V_D)=e^2(n_T+n_B)/C_{BG}+E_F(n_B)
\end{equation}
Here $E_F(n)=sgn(n)\hbar v_F\sqrt{\pi |n|}$ is the Fermi energy relative to the charge neutrality point in monolayer graphene at a carrier density $n$;
$sgn$ represents the sign function. Similarly, the bottom layer Fermi energy is the sum of the electrostatic potential difference between the
layers and the Fermi energy of the top layer:
\begin{equation}
E_F(n_B)=e^2n_T/C_{int}+E_F(n_T)
\end{equation}
where $C_{int}$ is the interlayer capacitance. Using Eqs. (1, 2) and $C_{int}$ as a fitting parameter, we calculate $n_T$ and $n_B$ as a
function of $V_{BG}$. An excellent fit to the experimental data is obtained for $C_{int}=6.9$ $\mu$F$\cdot$cm$^{-2}$ [solid lines in Fig. 3(c)].
Remarkably, this value is in good agreement with the inter-layer capacitance expected theoretically for a Bernal stacked bilayer,\cite{min} suggesting that
the separation of the two layers in twisted bilayer graphene is close to that of a Bernal stacked bilayer.
Two previous experimental studies \cite{schmidt,sanchez} which examined electron transport in twisted bilayer graphene consisting of two
exfoliated graphene mono-layers reported $C_{int}$ values of 0.6 $\mu$F$\cdot$cm$^{-2}$,\cite{schmidt} and 6.8 $\mu$F$\cdot$cm$^{-2}$.\cite{sanchez}

Figure 4 shows the $\rho_{xx}$ contour plot as a function of $V_{BG}$ and $B$ probed in the twisted bilayer sample of Fig. 3.
The charge neutrality (Dirac) point is reached at back-gate bias $V_D=68$ V. The data show a QHS pattern which stems from the
QHSs superposition of the two decoupled monolayers. To map the position of the observed QHSs, we use Eqs. (1) and (2) to calculate the layers densities
as a function of $V_{BG}$ and $B$, with the only difference that the Fermi energy depends on both density and magnetic field as $E_F=E_N$, where
$E_N=sgn(N)v_F \sqrt{2e\hbar B|N|}$ is the energy of the $N^{th}$ LL in monolayer graphene, and $N=Int[nh/4eB]$ is the LL index; $Int$
is the nearest integer function. Using $C_{int}=6.9$ $\mu$F$\cdot$cm$^{-2}$ extracted from Fig. 3 data analysis, we calculate $n_B$ and $n_T$
at fixed $B$ and $V_{BG}$ values, which are then converted into layer filling factors $\nu_{T,B}=n_{T,B}h/eB$.
The black (red) lines in Fig. 4 represent the calculated position of half-filled LLs, i.e. $\nu_{B,T}=\pm0,4,8,12...$ for the bottom (top) layer.
The $\rho_{xx}$ maxima are in excellent agreement with the calculations, quantitatively confirming that the QHS sequence of twisted bilayer
graphene is a superposition of the QHSs of the two graphene monolayers.

In summary, using a combination of Raman spectroscopy and magnetotransport measurements we established that CVD-grown bilayer graphene on Cu
consists of a mixture of Bernal stacked and twisted monolayer domains. The Bernal stacked domains show QHSs at filling factors $\nu=4,8,12$,
in agreement with data in exfoliated bilayer graphene. The twisted bilayer graphene domains display a superposition of the individual QHSs of
two grapehene monolayers, which allows us to extract the layer densities and inter-layer capacitance. The layer stacking determined from
magnetotransport data correlates with the FWHM of the Raman 2D band.

This work was supported by ONR, NRI-SWAN, and the W.M. Keck Foundation. Part of this work was performed at the National High Magnetic Field Laboratory,
which is supported by NSF (DMR-0654118), the State of Florida, and the DOE.

\end{document}